\documentclass[prl,aps,twocolumn,floats,showpacs,epsfig]{revtex4}

\usepackage{epsfig}

\newcommand{\be}{\begin{equation}}
\newcommand{\ee}{\end{equation}}
\newcommand{\bea}{\begin{eqnarray}}
\newcommand{\eea}{\end{eqnarray}}

\newcommand{\p}{\partial}

\newcommand{\rd}{\mbox{d}}
\newcommand{\ri}{\mbox{i}}
\newcommand{\re}{\mbox{e}}

\begin{document}
\title{Isospin excitations of a trapped 1D gas of attractively interacting fermions}

\author{M. Colom\' e-Tatch\'e $^1$,  G. V. Shlyapnikov$^{1,2}$ and A.  M. Tsvelik$^3$}
\affiliation{\mbox{$^1$ Laboratoire de Physique Th\'eorique et Mod\'eles Statistiques, Universit\'e. Paris Sud, CNRS, 91405~Orsay, France} \\
\mbox{$^2$ Van der Waals-Zeeman Institute, University of Amsterdam, Valckenterstraat 65/67, 1018 XE Amsterdam, The Netherlands}\\
\mbox{$^3$Department of  Condensed Matter Physics and Materials Science, Brookhaven National Laboratory, Upton, NY 11973-5000, USA}}
\date{\today}

\begin{abstract}
We consider a gas of fermions with a short-range attractive intercomponent interaction in a parabolic external potential
and derive the conditions of the local density approximation. The obtained spectrum of quasiparticle (isospin) excitations 
shows equidistant low-energy levels, which is equivalent to a linear momentum dependence and is fundamentally different 
from the ordinary Dirac spectrum in the spatially uniform case.


\end{abstract}

\maketitle


Fast progress in experiments with cold atoms has led to the creation of one-dimensional
(1D) atomic gases by (tightly) confining the motion of atoms in two directions to zero point
oscillations.  One-dimensional quantum gases show a remarkable
physics not encountered in higher dimensions. In particular, since the density of states in 1D increases 
towards the zone boundary, the effective strength of interactions in the 1D gas increases  with decreasing density. 
Therefore, decreasing density in a gas of attractively interacting fermions one can crossover from the BCS-like regime 
of strongly overlapped pairs to the regime of compact pairs forming a Bose-Einstein condensate (BEC). 
For spatially uniform systems this problem is well understood due to availability of exact solutions obtained 
by the Bethe Ansatz \cite{gaudin,yang,Essler}, in combination with powerful bosonization techniques 
\cite{Tsvelik1,Tsvelik2,Giamarchi}.

The 1D gases are usually obtained in an external harmonic potential, which introduces a finite size of the system
and makes it spatially nonuniform. Trapped 1D Bose gases have been intensively studied in the last years 
(see \cite{ PGS} for review). Recently, the 1D regime has been achieved for atomic fermions \cite{Esslinger},
and the discussion of trapped 1D Fermi gases was focused on the occurrence of the BCS-BEC crossover, manifestation of 
spin-charge separation, role of the imbalance between atomic components (see \cite{Trento,Kollath} for review). 
One of the key problems is revealing the influence of an external harmonic potential on the 
quasiparticle spectrum of this system.

In this paper we consider a 1D $N$-component Fermi gas with a point-like attraction in a parabolic trap. We first treat this 
problem in the limit of $N>> 1$ using the standard $1/N$-expansion \cite{Koberle}. In the limit of $N \rightarrow \infty$ the 
saddle point approximation becomes exact and one can deal with the spatially non-uniform distribution of particles
in a controllable way. We show that the spatial inhomogeneity strongly affects the quantization rules for the quasiparticle spectrum. As a result, 
the system in a parabolic trap cannot be mapped onto an integrable system in a rectangular box. We then discuss the applicability of the 
large-$N$ results for the experimentally relevant case of a two-component Fermi gas ($N=2$).  

In terms of field operators $\psi_j^+$ and $\psi_j$, with the index $j$ labelling the fermionic species, the bare Hamiltonian is: 
\begin{eqnarray}       \label{H}
&&H=\sum_j\int dx\Big[ - \frac{1}{2m}\psi^+_j\p_x^2\psi_j -\frac{m\omega^2x^2}{2}\psi_j^+\psi_j  \nonumber \\ 
&&-\frac{g}{N}\sum_{j^{\prime}\neq j}\psi_j^+\psi^+_{j^{\prime}}\psi_{j^{\prime}}\psi_j\Big],
\end{eqnarray}
where $m$ is the atom mass, $\omega$ is the trap frequency, the coupling constant of the point-like attraction is written as
$-g/N$ with $g>0$, and we set $\hbar=1$. We will work in the thermodynamic limit defined by the relations 
\bea
E_F=\omega {\cal N}/N = \mbox{const}, ~~ {\cal N} \rightarrow \infty, ~~ \omega \rightarrow 0;
\eea
with ${\cal N}$ being the total number of particles, and $E_F$ the Fermi energy.

We first briefly outline the results for the uniform case where the system is in a rectangular box with periodic boundary conditions. 
Then, in the thermodynamic limit the excitation spectrum consists of a gapless collective mode describing fluctuations of the total density (charge mode) 
and $N-1$ branches of particles (isospin modes) \cite{Schroer,Andrei}. The latters have spectral gaps: the particle of the $q$-th branch transforms according to the fundamental 
representation of the SU(N) group with the Young tableau consisting of one column of height $q$. In the limit of $\gamma << 1$ where 
\be
\gamma =\frac{mg}{\pi n} \label{gamma}
\ee
and $n$ is the mean density of each fermionic component, the spectrum of the gaped particles is approximately Lorentz invariant: 
\bea
&& \epsilon_q(p) =\sqrt{(pv_F)^2 + M_q^2}, \label{spectrum}\\
&& M_q = M_0\sin(2\pi q/N), ~~ q=1,...N-1;\nonumber\\
&& M_0 = CE_F\gamma^{1/N}\exp(-2\pi/\gamma), \label{gap}
\eea 
where $C$ is a numerical constant. At a finite $N$ these particles interact and the ones with $q > 1$ are bound states of the fundamental particle with $q=1$. 
However, in the limit of $N \rightarrow \infty$ the interaction vanishes and the only  particles which remain are the particle with $q=1$ and its antiparticle 
having $q= N-1$. The case of $N >> 1$ can be treated perturbatively using $1/N$ expansion \cite{Koberle}.

This approach can be employed for a spatially nonuniform system as is done in this paper. To make calculations easier we assume that in the major part of the 
trap the coupling constant (\ref{gamma}) in which $n$ now depends on the coordinate, is small. This allows us to use the Thomas-Fermi density profile 
for non interacting fermions:
\begin{equation}  \label{TF}
\!\!n(x)\!=\!n_0\sqrt{1 - x^2/x_0^2};\, {\cal N}\!\!=\!\frac{\pi}{2}n_0x_0N;\, x_0\!\!=\!\!\sqrt{\frac{2{\cal N}}{mN\omega^2}},
\end{equation}
where $x_0$ is the Thomas-Fermi (half)size of the sample. 
The Thomas-Fermi approximation breaks down at distances near the Thomas-Fermi boundary, where $dn^{-1}(x)/dx\sim 1$ and, hence,
$(1-x^2/x_0^2)^{1/2} \sim {\cal N}^{-1/3}$. 
The condition of weak coupling breaks down in a narrow range of distances near the boundary, where $\gamma(n(x))\sim 1$ 
and $(1 - x^2/x_0^2)^{1/2} \sim \gamma(0) \equiv mg/\pi n_0\ll 1$. 

Assuming that the weak coupling approximation holds, we can linearize the 
spectrum introducing right and left moving fermions $R,L$:
\begin{eqnarray}  
\psi_j(x) = \re^{-\ri k_F(x)x}R_j(x) + \re^{\ri k_F(x) x}L_j(x). \nonumber
\end{eqnarray}
The coordinate-dependent Fermi momentum is $k_F(x) = \pi n(x)$, and Hamiltonian (\ref{H}) takes the form:
\begin{eqnarray}    \label{GN}
&& H=\sum_j\int_{-\tau_0}^{\tau_0}d\tau\Big[\ri(- R_j^+\p_{\tau}R_j+L_j^+\p_{\tau}L_j)  \nonumber \\
&& -\frac{\gamma(\tau)}{2N}(R^+_jL_j)\sum_{j'\neq j}(L^+_{j'} R_{j'})\Big],
\end{eqnarray}
where $\gamma$ is given by Eq.~(\ref{gamma}), and
\begin{equation}   \label{tau}
\tau = \int_0^x dx^{\prime}/v_F(x^{\prime})=\omega^{-1}\arcsin(x/x_0),
\end{equation}
with the Fermi velocity $v_F=k_F/m$, and $\tau_0 = \pi/2\omega$.

At $N >> 1$ the interaction can be treated in the saddle-point approximation which becomes exact in the 
limit of $N \rightarrow \infty, \gamma =$ const. To develop such an approximation one does the Hubbard-Stratonovich 
transformation introducing an auxiliary complex field $\Delta(\tau,t)$, and formally integrates over the fermions. 
The resulting effective action is
\bea
\!S\!\!=\!\!N\int\!\!dtd\tau\!\left\{\frac{|\Delta(t,\tau)|^2}{2\gamma(\tau)}\!-\!\mbox{Tr}\ln\!
\left[\!
\begin{array}{cc}
\ri(\p_t-\p_{\tau}) & \Delta\\
\Delta^* & \ri(\p_t + \p_{\tau})
\end{array}
\right]\right\},  \nonumber
\eea
where $t$ is the Matsubara time. Expanding the logarithm in gradients of $\Delta$ we obtain the Lagrangian density:
\begin{widetext}
 \bea
{\cal L} = \frac{N}{4\pi}\left\{\left[\frac{2\pi}{\gamma(\tau)} - \ln(\Lambda/|\Delta|)\right]|\Delta|^2 + \frac{1}{2}|\Delta|^{-2}\left[|\p_t\Delta|^2 + |\p_{\tau}\Delta|^2\right]\right\} + ... \label{dens}
\eea
\end{widetext}
where $\Lambda\sim E_F$ is the high-energy cut-off, and the dots stand for the terms containing higher powers  
of $\p_a\Delta^{-1}$ ($a = t, \tau$). These terms are small under the condition $|\Delta_0|\equiv|\Delta(0)|\gg\omega$, since the characteristic scale of $\tau$ is 
$\omega^{-1}$. This is equivalent to having the system size $x_0$ much larger than the largest correlation length $\xi_0 =v_F(0)/|\Delta_0|$.  

We now write $\Delta$ in the form 
\be
\Delta = \rho\exp(\ri\phi) \label{OP}
\ee
and obtain the action in which fluctuations of the phase $\phi$ are decoupled from the fluctuations of the amplitude $|\Delta| \equiv \rho$:
\bea
&& {\cal L} = {\cal L}[\phi] + {\cal L}[\rho] \label{L},\\
&& {\cal L}[\phi] =\frac{N}{8\pi}\left[(\p_t\phi)^2 + (\p_{\tau}\phi)^2\right] \label{phi},\\
&& {\cal L}[\rho] =\frac{N}{4\pi}\Big\{\rho^2\left[\frac{2\pi}{\gamma(\tau)} - \ln(\Lambda/\rho)\right]  \nonumber  \\ 
&& + \frac{1}{2}\left[\frac{(\p_t\rho)^2}{\rho^2} + \frac{(\p_{\tau}\rho)^2}{\rho^2}\right] +...\Big\}. \label{rho}
\eea
There are two features of this expansion, with a different level of robustness under deviations from the condition of weak coupling, $\gamma << 1$. 
First of all, the phase $\phi$ is decoupled from the amplitude field, which is a simple consequence of $\phi$ being the Goldstone mode of the 
Charge Density Wave order parameter field (\ref{OP}). As such it should be gapless and weakly coupled to other excitations.    
However, the second feature, namely the fact that the stiffness of the field $\phi$ is independent of the particle density $n(x)$, 
holds only in the limit of $\gamma << 1$. As follows from the Bethe Ansatz calculations valid in 
the uniform case, the stiffness (or the Luttinger parameter $K_c$) starts to acquire the density dependence at $\gamma \sim 1$.  

Since the action following from Eqs.~(\ref{L})-(\ref{rho}) is proportional to large $N$, fluctuations are suppressed. In particular, one can find 
excitation energies by considering the Dirac Hamiltonian with the coordinate-dependent mass determined by the minimum of the action (\ref{rho}):
\bea
2\rho[2\pi/\gamma(\tau) - \ln(\Lambda/\rho)] - \rho^{-1}\p_{\tau\tau}\ln\rho + ... = 0 \label{min}
\eea
The solution can be represented as 
\bea
\rho(\tau)\!=\!\rho(0)\exp\left\{2\pi^2[n(0)\!-\!n(\tau)]/gm\!+\!\frac{1}{2\rho^2}\p_{\tau\tau}\ln\rho +...\right\}. \nonumber
\eea
The second term in the exponent of this expression can be omitted if one satisfies the inequality
\be        \label{cond}
\Delta_0/\omega\gg 1/\sqrt{\gamma_0},
\ee
where $\gamma_0=\gamma(0)$. One then has
\bea
\rho(\tau) =\Delta_0\exp\{2\pi^2[n_0-n(\tau)]/mg\}.  \label{local}
\eea

Now we find the excitation spectrum of quasiparticles. As we have said, in the limit of $N \rightarrow \infty$ their spectrum is decoupled from 
the phase excitations. Then the fermions are described by the effective Dirac Hamiltonian
\bea
H_F = \sum_j\int \rd\tau (R_j^+,L^+_j)\left(
\begin{array}{cc}
-\ri \frac{\rd}{\rd\tau} & \rho(\tau)\\
\rho(\tau) & \ri\frac{\rd}{\rd\tau}
\end{array}
\right)
\left(\begin{array}{c}
R_j\\
L_j
\end{array}
\right),\label{Dirac}
\eea
where $\rho(\tau)$ is given by Eq.~(\ref{local}). A uniform system in a rectangular box with periodic boundary conditions,
is described by the same Hamiltonian, with $\rho(\tau)=\Delta_0$. Equations of motion following from this Hamiltonian yield 
solutions in the form of plane waves with the spectrum
\bea
\epsilon_k^2 =(\pi kv_F/l)^2 + \Delta_0^2, ~~ k =0,\pm 1, \pm 2 ...,  \label{rb}
\eea
where $2l$ is the length of the box. This corresponds to the spectrum of $q=1, N-1$ excitation branches in Eq.(\ref{spectrum}).  

Actually, having derived Eq.~(\ref{local}) for $\rho(\tau)$ entering the Hamiltonian $H_F$ (\ref{Dirac}), we obtained the local
density approximation for our nonuniform system. Our results show that in the Dirac Hamiltonian (\ref{Dirac}) one can use
the same expression for the mass (gap) as in the uniform case, but the exponent is coordinate dependent through the spatial dependence of the
density (parameter $\gamma$) and the coordinate dependence of the preexponential factor is omitted. However, the quasiparticle spectrum is 
quite different from that of Eq.~(\ref{rb}).

Eigenfunctions  of Hamiltonian (\ref{Dirac}) can be written as  
\bea
R = \rho^{1/2} F, ~~ L = (\epsilon R + \ri R_{\tau})/\rho,
\eea
where the function $F(\tau)$ satisfies the equation
\bea   \label{Sch}
\!-F_{\tau\tau}\!+\!\Big[\rho^2(\tau)\!-\!\epsilon^2\!+\!\Big(\frac{\rho_{\tau\tau}}{2\rho}\!-\!\left(\frac{\rho_{\tau}}{2\rho}\right)^2\!+\!\ri\epsilon\frac{\rho_{\tau}}{\rho}\Big)\Big]F\!=\!0,\!
\eea
and the notations $F_{\tau}$ and $F_{\tau\tau}$ mean the first and second derivative with respect to $\tau$. 
For finding the lowest eigenstates one can approximate 
\be   \label{rhoapprox}
\rho^2(\tau)\approx \Delta_0^2\left[1 + \frac{2\pi\omega^2\tau^2}{\gamma(0)}\right]
\ee
and check that under the condition (\ref{cond}) the terms in the round brackets in Eq.~(\ref{Sch}) are small at least as 
$(\omega/\sqrt{\gamma_0}\Delta_0)^{1/2}$ compared to both terms on the rhs of Eq.~(\ref{rhoapprox}). Then, omitting the terms in the round brackets, 
for the energy eigenvalues we obtain:
\be   \label{Eklow}
\epsilon_k = \sqrt{\Delta_0^2 + \Delta_0\omega(8\pi/\gamma_0)^{1/2}( k + 1/2)},
\ee
where $k >0$ is an integer. Identifying $k$ as a (rescaled) momentum of a particle confined in a rectangular box and comparing 
Eq.~(\ref{Eklow}) with Eq.~(\ref{rb}) one sees that the quasiparticle energy levels in the parabolic trap cannot be mapped onto the Dirac particle spectrum in the box.

For finding high-energy eigenstates one can use the WKB quantization rule. Omitting again the terms in the round brackets in Eq.~(\ref{Sch}) we have
\bea  \label{WKB}
\int_{-\tau(\epsilon)}^{\tau(\epsilon)} d\tau \sqrt{\epsilon^2 -\rho^2(\tau)}=\pi k,
\eea
where $\tau(\epsilon)$ is determined by the condition $\epsilon=\rho(\tau)$. 
Rewriting Eq.~(\ref{local}) as
$$\rho(\tau)=\Delta_0\exp\{2\pi(1-\cos\omega\tau)/\gamma_0\}$$
we notice that considering $\epsilon\ll E_F$ one can write the exponent of $\rho(\tau)$ as $\pi\omega^2\tau^2/\gamma_0$. 
This leads to 
\begin{equation}    \label{tauE}
\tau(\epsilon)\approx\frac{1}{\omega}\sqrt{\frac{\gamma_0}{\pi}\ln{(\epsilon/\Delta_0)}}.
\end{equation} 
Then performing the integration in Eq.~(\ref{WKB}) we obtain:
\begin{equation}     \label{WKBdisp}
\frac{\sqrt{\epsilon_k^2-\Delta_0^2}}{\omega}\left(\frac{2\gamma_0}{\pi}\right)^{1/2}\!\!\sqrt{\ln\left(1+\beta\frac{\epsilon_k^2-\Delta_0^2}{\Delta_0^2}\right)}\!=\!\pi k,
\end{equation}
where for eigenstates near the bottom of the gap, that is for $(\epsilon_k-\Delta_0)\ll\Delta_0$, the coefficient $\beta$ is equal to $\pi^2/16$ and Eq.~(\ref{WKBdisp}) gives
the same result as Eq.~(\ref{Eklow}) at large quantum numbers $k$: $\epsilon_k^2=\Delta_0^2+\Delta_0\omega(8\pi/\gamma_0)^{1/2}k$.
For excitation energies $\epsilon_k\gtrsim\Delta_0$ one has $\beta$ close to unity, and 
in the limit of $\epsilon_k\gg\Delta_0$ corresponding to quantum numbers $k\gg\Delta_0\sqrt{\gamma_0}/\omega$, Eq.~(\ref{WKBdisp}) yields:
\be        \label{Ekhigh}
\epsilon_k =\left(\frac{\pi^3}{4\gamma_0}\right)^{1/2}\frac{\omega k}{\sqrt{\ln[(\pi^3/2\gamma_0)^{1/2}\omega k/\Delta_0]}}.
\ee
As we see, this result also differs significantly from the simple Dirac spectrum.

In Fig.~1 we compare the results of Eqs.~(\ref{Eklow}) and (\ref{WKBdisp}) with numerical calculation of the spectrum
from Eq.~(\ref{Sch}). One sees a good agreement even for $\Delta_0/\omega$ about 20,
and for larger values of $\Delta_0/\omega$ the agreement is nearly perfect.  
\begin{figure}[t]
\centerline{
        \mbox{\includegraphics[width=3.0in]{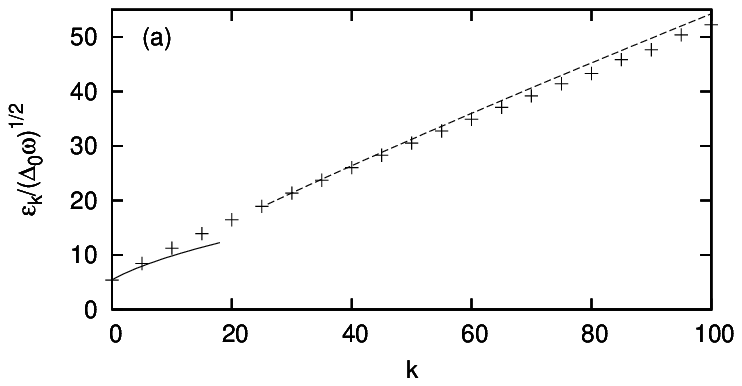}}
 }
\vspace{-1.2cm}
\centerline{
        \mbox{\includegraphics[width=3.0in]{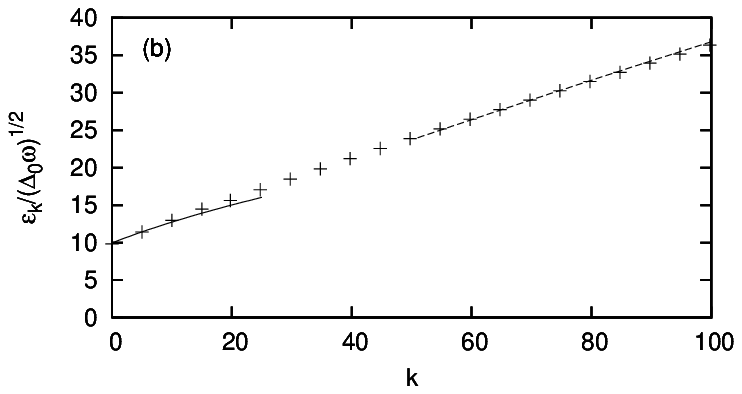}}
        }
  \caption{Energy levels for: (a) $\Delta_0 /\omega=26$ ($\gamma_0=0.44$); (b) $\Delta_0 /\omega=95$ ($\gamma_0=0.5$).
The solid and dashed curves show the results of Eq.~(\ref{Eklow}) and the results of 
Eq.~(\ref{WKBdisp}) with $\beta=1$. The crosses are the 
results of numerical solution of Eq.~(\ref{Sch}).}
  \label{Fig.1}
\end{figure}

We believe that our main result, a pronounced difference between quantization rules for the quasiparticle spectrum in harmonic and 
rectangular confining potentials, remains valid in the experimentally relevant case of a two-component Fermi gas ($N=2$). 
It is likely that the local density approximation for the mass gap $\rho$ calculated along the lines of \cite{Krivnov,Larkin}, 
together with Eq.~(\ref{WKB}), remain  robust. The main difference of small from large $N$ is a dramatic increase of 
interactions between quasiparticles. This, however, will not have a strong effect on the energy levels as soon as the number of 
excited quasiparticles is macroscopically small. The interactions affect commutation relations of the quasiparticle creation and 
annihilation operators (see, e.g., Chapter 34 of \cite{Tsvelik2} and references therein) and this leads to changes 
in the behaviour of matrix elements and correlation functions \cite{EssTs}. 


For a trapped 1D atomic ultracold Fermi gas one can expect the number of particles ${\cal N}\sim 10^4$. Then, recalling that in the two-species gas the Fermi energy
is $E_F={\cal N}\omega/2$, for $\gamma_0\approx 1$ we obtain $\Delta_0/\omega\approx 10$ and our results are applicable.
The isospin modes can be excited optically, for example by using pulses of polarised $\sigma^-$ light acting on one of the atomic components (spin-up)
and pulses of $\sigma^+$ light acting on the other component (spin-down). The $\sigma^-$ and $\sigma^+$ pulses provide periodic optical potentials
shifted by a quarter of a wavelength with respect to each other, so that the minimum of the $\sigma^-$ potential corresponds to the maximum of the
$\sigma^+$ potential \cite{dalibard} and the sum of the two potentials is zero. Thus, the spin-up and spin-down particles get kicks in the opposite directions,
which creates isospin modes. At the same time, charge (density) modes corresponding to in-phase oscillations of the two components are not excited.    

In conclusion, we have found the quasiparticle (isospin) spectrum for attractively interacting fermions in a parabolic potential. 
The spectrum shows equidistant low-energy levels (linear momentum dependence) and is drastically different from the ordinary 
Dirac spectrum in the spatially uniform case. Experimental verification of this result will provide a clear demonstration
of the fact that the parabolic confinement can fundamentally change the properties of the quantum gas.

We are grateful to B. L. Altshuler, I. L. Aleiner, J. Dalibard, F. Gerbier, D.L. Kovrizhin, and L.P. Pitaevskii for discussions 
and acknowledge support of Institut Henri Poincar\'e 
during the workshop "Quantum Gases" where part of this work has been done.  
The work was also supported by the IFRAF Institute, by
ANR (grants 05-BLAN-0205 and 06-NANO-014-01), by the QUDEDIS program of ESF, and by the Dutch Foundation FOM.  
AMT was  supported  by US DOE under contract number DE-AC02 -98 CH 10886. 
LPTMS is a mixed research unit No. 8626 of CNRS and Universit\'e Paris Sud.  


\begin{thebibliography}{99}


\bibitem{gaudin} M. Gaudin, Phys. Lett. A{\bf 24}, 55 (1967). 
\bibitem{yang} C. N. Yang, Phys. Rev. Lett. {\bf 19}, 1312 (1967).
\bibitem{Essler}F. H. L. Essler, H. Frahm, F. Goehmann, A. Kluemper, and V. E. Korepin,  
{\it The One-Dimensional Hubbard Model} (Cambridge University Press, Cambridge, UK, 2005).
\bibitem{Tsvelik1} A. O. Gogolin, A. A. Nersesyan and A. M. Tsvelik, {\it
Bosonization and Strongly Correlated Systems} (Cambridge University Press, Cambridge 1999). 
\bibitem{Tsvelik2} A. M. Tsvelik {\it Quantum Field Theory in Condensed Matter Systems} (Cambridge University Press, Cambridge 2003). 
\bibitem{Giamarchi} T. Giamarchi, {\it Quantum Physics in One Dimension} (Oxford University Press, Oxford, 2004).
\bibitem{PGS} D.S. Petrov, D.M. Gangardt, and G.V. Shlyapnikov, J. Phys. IV (France) {\bf 116}, 5 (2004);
Y. Castin, {\it ibid} {\bf 116}, 89 (2004). 
\bibitem{Esslinger} H. Moritz {\it et al}, Phys. Rev. Lett. {\bf 94}, 210401 (2005).
\bibitem{Trento} S. Giorgini, L.P. Pitaevskii, and S. Stringari, arXiv:0706.3360.
\bibitem{Kollath} C. Kollath and U. Schollw\"ock, New Journal of Physics, {\bf 8}, 220 (2006).
\bibitem{Koberle} R. Koberle, V. Kurak and J. A. Swieca, Phys. Rev. D {\bf 20}, 897 (1979).
\bibitem{Schroer} B. Schroer, T. T. Truong and P. Weisz, Phys. Lett. B {\bf 63}, 422 (1976). 
\bibitem{Andrei} N. Andrei, Phys. Lett. B {\bf 90}, 106 (1980).
\bibitem{Krivnov} V.Ya. Krivnov and A.A. Ovchinnikov, Zh. Eksp. Teor. Fiz. {\bf 67}, 1568 (1974)
[Sov. Phys. JETP {\bf 40}, 781 (1975)].
\bibitem{Larkin} A.I. Larkin and J. Sak, Phys. Rev. Lett. {\bf 39}, 1025 (1977). 
\bibitem{EssTs} F. H. L. Essler and A. M. Tsvelik, Phys. Rev. Lett {\bf 90}, 126401 (2003).
\bibitem{dalibard} This scheme lies in the basis of laser cooling below the Doppler limit:
J. Dalibard and C. Gohen-Tannoudji, J. Opt. Soc. Am. {\bf 6}, 2023 (1989). A similar scheme was used
for controlled coherent transport in spin-dependent optical lattice potentials: O. Mandel {\it et al},
Phys. Rev. Lett. {\bf 91}, 010407 (2003).

\end{thebibliography}
\end{document}